# New Electronic Phase Transitions in α-(BEDT-TTF)$_2$KHg(SCN)$_4$

M. V. Kartsovnik[1§], D. Andres[1], W. Biberacher[1], P. D. Grigoriev[2], E. A. Schuberth[1] and H. Müller[3]

[1] *Walther-Meissner-Institute, Bavarian Academy of Sciences, Walther-Meissner-Str. 8, D-85748 Garching, Germany; e-mail:*
[2] *L. D. Landau Institute for Theoretical Physics, 142432 Chernogolovka, Russia;*
[3] *European Synchrotron Radiation Facility, F-38043 Grenoble, France*

**Abstract.** α-(BEDT-TTF)$_2$KHg(SCN)$_4$ is considered to be in the charge-density-wave (CDW) state below 8 K. We present new magnetoresistance data suggesting that the material undergoes a series of field-induced CDW (FICDW) transitions at pressures slightly exceeding the critical pressure $P_c$ at which the zero-field CDW state is destroyed. Further, we argue that a novel kind of FICDW transitions, entirely determined by a superposition of the strong Pauli and quantizing orbital effects of magnetic field on the CDW wavevector, arises when the field is strongly tilted towards the conducting layers. These new transitions can take place even in the case of a relatively well nested Fermi surface. Finally we report on the superconducting (SC) state and its coexistence with the CDW in the title compound under quasi-hydrostatic pressure. Below $P_c$ the material is most likely a heterogeneous SC/CDW mixture, with the SC phase persisting down to ambient pressure. The SC onset temperature appears to drastically increase upon entering the SC/CDW coexistence region.

## 1. INTRODUCTION

The conducting system of α-(BEDT-TTF)$_2$*M*Hg(SCN)$_4$ (where *M* = K, Tl, Rb or NH$_4$) includes both the quasi-one-dimensional (q1D) electron-like and quasi-two-dimensional (q2D) hole-like carriers [1]. The salts with *M* = K, Tl and Rb undergo a phase transition at $T_p \cong$ 8-10 K which is associated with the nesting of the q1D Fermi surface (FS). The 2$k_F$-superstructure, observed originally in the angle-dependent magnetoresistance experiments (see e.g. [2] for a review), was recently confirmed by X-ray studies [3], thus suggesting a charge-density-wave (CDW) formation in these compounds. Current-voltage characteristics taken on the *M*=K salt below $T_p$ [4] reveal the existence of a threshold electric field whose behaviour can be interpreted in terms of a CDW with a *d*-wave order parameter [5].

It is important that the transition temperature $T_p$ and, therefore, the relevant energy gap are much smaller than it is usually met in CDW materials. This leads to a very strong influence of a magnetic field on electronic properties, giving rise to numerous anomalies which stimulated high interest to these compounds for over a decade. The "magnetic field – temperature" (*B*–*T*) phase diagram studied extensively in the last few years [6-8] is in very good agreement with theoretical predictions for a CDW system [9-11], providing a strong argument for the CDW origin of the electronic instability. At ambient pressure, it is dominated by coupling of spins of the interacting electrons to the field, the so-called *Pauli effect*, which suppresses the CDW in a way analogous to the paramagnetic suppression of a singlet superconductivity. In particular, one of the most prominent anomalies, the so-called *kink* transition, is associated with the transformation of the zero-field $CDW_0$ state to a high-field $CDW_x$ state which is analogous to the Larkin-Ovchinnikov-Fulde-Ferrel state in superconductors.

Under pressure the nesting property of the q1D FS becomes worse. This can be described by an enhancement of the second-order interchain transfer integral $t'_\perp$ corresponding to the next-nearest-chain transfer in the plane of conducting layers. This obviously leads to a decrease of the zero-field transition temperature $T_p(B=0)$. At these conditions, the second, *orbital* effect becomes important, leading to an increase of $T_p$ with magnetic field [12]. At a high enough pressure, $P_c \cong 2.3\text{-}2.5$, the CDW is completely suppressed at $B=0$ but, according to theory [10,13], it can be restored in the form of a cascade of quantized field-induced CDW (FICDW) phases. This is similar to the well known field-induced spin-density-wave (FISDW) phenomenon (see [14,15] for a review). However, in our case one should take into account a competition between the orbital and Pauli effects. In the next section, we present new data on the magnetoresistance of $\alpha$-(BEDT-TTF)$_2$KHg(SCN)$_4$ under quasi-hydrostatic pressure which support the existence of FICDW in this material. Further, we argue that a novel manifestation of the orbital quantization which originates from *simultaneous* effects of Pauli and orbital coupling of a high field to a CDW is observed in our compound already at ambient pressure in fields strongly tilted towards the conducting layers.

In the last section we consider the superconducting (SC) state in $\alpha$-(BEDT-TTF)$_2$KHg(SCN)$_4$ and its coexistence with the CDW state under quasi-hydrostatic pressure. It was recently shown that the SC state emerges at $T_c \leq 0.1$ K under quasi-hydrostatic pressure above $P_c$, when the zero-field CDW is completely suppressed [16]. The transition is rather sharp but the maximum $T_c$ is more than an order of magnitude lower than that found previously under uniaxial strain [17,18]. Nevertheless, the superconductivity does not disappear completely at lowering the pressure below $P_c$ and small fractions of the superconducting phase persist down to $P = 0$. Moreover, the onset of superconductivity appears to drastically increase up to ~ 0.3 K upon entering the CDW region of the phase diagram.

## 2. FIELD-INDUCED CHARGE-DENSITY-WAVE TRANSITIONS

### 2.1. Magnetoresistance in perpendicular fields, under pressure

Figure 1 shows interlayer magnetoresistance of $\alpha$-(BEDT-TTF)$_2$KHg(SCN)$_4$ in fields perpendicular to the conducting layers recorded at $T = 0.1$ K at different pressures. Strong Shubnikov-de Haas (SdH) oscillations with the fundamental frequency of 670 to 740 T (depending on pressure) originate from the q2D band which remains metallic in the CDW state. At low pressure these oscillations are superimposed on a very high smooth background typical of the CDW state of this compound. With increasing $P$ the CDW is gradually suppressed which is, in particular, reflected in a rapid decrease of the background magnetoresistance. Further, at $P \geq P_c$, new oscillatory features emerge in the $R(B)$ curves. They are most prominent at $P = 3.5\pm0.5$ kbar and fade away outside this pressure interval. It is important that this interval exactly matches the conditions at which the FICDW transitions are expected for our compound [10,12]: it corresponds to the antinesting parameter $t'_\perp$ exceeding but still close to the critical value $t'^*$ at which the zero-field CDW vanishes.

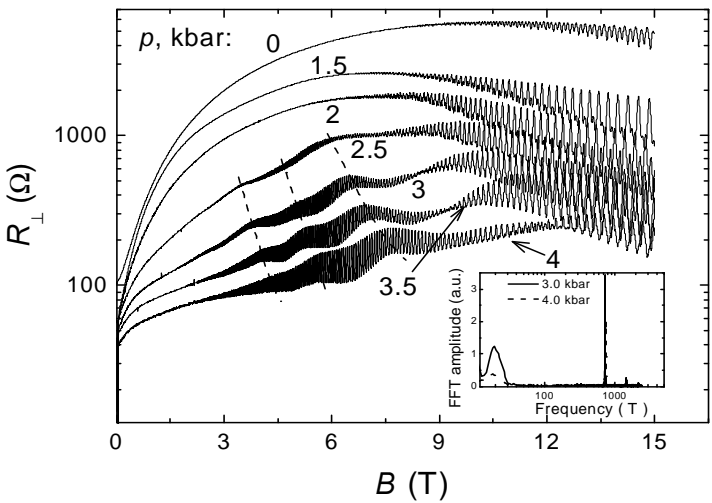

Figure 1. Interlayer magnetoresistance of $\alpha$-(BEDT-TTF)$_2$KHg(SCN)$_4$ in perpendicular fields at different pressures; $T = 100$ mK. Inset shows fast Fourier transformation (FFT) spectra of the curves corresponding to P = 3.0 and 4.0 kbar.

The new features are approximately periodic in $1/B$ with the frequency $F \approx 20$ T

which does not change significantly between 2.5 and 4 kbar. However, the positions of the features shift notably to higher fields with increasing pressure (see dashed lines in Fig. 1), in line with their proposed FICDW nature. Indeed, while the value of $t'$, determining the frequency, is expected to change only slightly within the narrow pressure range, already these small changes are supposed to significantly affect the $B$–$T$ phase diagram when $t'_\perp$ is close to $t'^*$.

From the frequency of the FICDW transitions one can roughly estimate the value of $t'$ at the pressures near critical: $t'_\perp \cong \frac{\pi}{8}\frac{e}{c}a_y v_F F = 0.55$ meV (here $e$ is the electron charge, $c$ is the light velocity, $a_y = 10^{-7}$ cm [1] is the interchain distance in the conducting plane, $v_F = 6.5 \times 10^6$ cm/s [19] is the Fermi velocity on the q1D FS). This is very close to the mean-field critical value [20] $t'^* = k_B T_{c0}/1.13 \approx 0.63$ meV, assuming that the transition temperature for a perfectly nested FS is equal to $T_c(P=0) = 8$ K.

## 2.2. Ambient-pressure FICDW transitions in strongly tilted fields

When the field is tilted by more than ~60° from the direction perpendicular to the layers, a series of features emerges in the resistivity and magnetisation recorded as a function of field [21-24]. The behaviour of the new structure can be described as follows. First, the features are independent of the azimuthal orientation of the field (see Fig. 2a) but depend on the tilt angle $\theta$, shifting to lower fields with increasing $\theta$. As shown in Fig. 2b, the shift rate increases as $\theta$ approaches 90°. It is therefore natural to suggest that the orbital effect (which is determined by the perpendicular field component $B\cos\theta$) is responsible for this behaviour. The interval of the tilt angles at which the structure is observed, 60° ≤ $\theta$ < 90°, suggests that the orbital effect should be finite but sufficiently small. Further, as seen in Fig. 2b, the new features arise only above the kink field $B_k$, i.e. when the CDW wave vector starts to vary with field due to the Pauli effect. Therefore, a possible impact of the strong Pauli effect must also be taken into consideration. Finally, a clear hysteresis between field sweeps up and down (Fig. 2a) is a likely evidence of discontinuous changes in the system. The following qualitative model [24] proposes the observed structure to be a manifestation of FICDW transitions originating from a superposition of the strong Pauli and orbital effects on the CDW state.

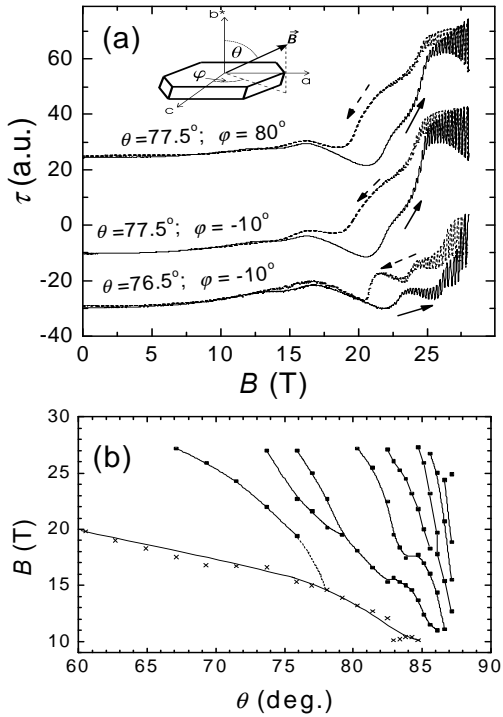

Figure 2. (a) Magnetic torque as a function of field at different orientations. Angles $\theta$ and $\varphi$ are defined in the inset. (b) Positions of the kink transition (crosses) and new high-field anomalies in the torque (circles) versus tilt angle.

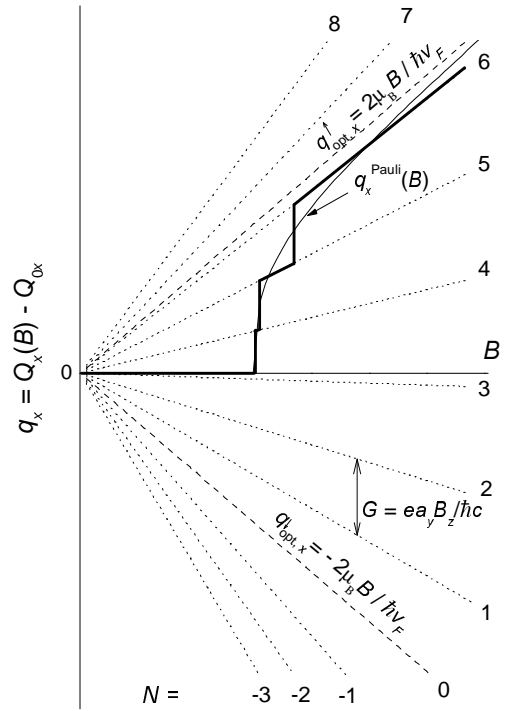

Figure 3. Schematic illustration of the superposition of the Pauli effect and orbital quantization on the CDW nesting vector (see text).

We consider the field dependence of the $Q_x$ component of the nesting vector in a CDW system with a moderately imperfect nesting ($t'_\perp < t^*$). At zero field, $Q_{0x} \approx 2k_F$ corresponds to the optimal nesting; the entire FS is gapped. At a finite magnetic field the degeneracy between the CDW's with different spin orientations is lifted. Treating each spin subband independently, one can express the optimal nesting conditions as $Q_{opt,x}(B) = Q_{0x} \pm 2\mu_B B/\hbar v_F$ where $\mu_B$ is Bohr magneton and the sign +/- stands for the spins parallel/antiparallel to the applied field. This splitting of the optimal nesting conditions is illustrated by dashed lines in Fig. 3. Nevertheless, both subbands remain fully gapped and the system as a whole maintains the constant nesting vector $Q_{0x}$ up to the critical field $B_k \sim \Delta(B=0)/2\mu_B$. Above $B_k$, $Q_{0x}$ is no more a good nesting vector as it leads to ungapped states in both subbands. As shown by Zanchi et al. [10], the CDW energy can be minimized in this case by introducing a field dependent term $q_x^{Pauli} = Q_x(B) - Q_{0x}$ which is schematically represented in Fig. 3 by the thin solid line asymptotically approaching the value $2\mu_B B/\hbar v_F$. This obviously improves the nesting conditions for one of the spin subbands (say, the spin-up subband) at the cost of an additional "unnesting" of the other (spin-down).

Now it is important to take into account that the spin-down subband becomes unnested at $B > B_k$ and therefore is subject to a strong orbital effect. The situation is analogous to that with a large antinesting term $t'_\perp \geq t^*$. One can therefore expect that, like in the "conventional" FISDW or FICDW case, an orbital quantization condition be set on the system. However, unlike in the FISDW case, the quantized levels are counted from ($Q_{0x} - 2\mu_B B/\hbar v_F$) rather than from $Q_{0x}$. The corresponding values $q_{xN} = -2\mu_B B/\hbar v_F + NG$ (where $G = ea_y B_\perp/\hbar c$, and $B_\perp = B\cos\theta$ is the field component perpendicular to the layers) are shown by dotted lines in Fig. 3.

As a result, the most favourable values of the nesting vector above $B_k$ are determined by intersections of the continuous curve $q_x^{Pauli}$ with the straight lines $q_{xN}$, i.e. by the superposition of the Pauli and quantum orbital effects. Thus, with changing the field we obtain a series of discontinuous transitions between CDW subphases characterized by different quantized values of the nesting vector as schematically shown by thick lines in Fig. 3.

The multiple FICDW transitions can be observed when the distance $G$ between the quantized levels is smaller than $\mu_B B/\hbar v_F$. This condition is obviously not fulfilled for $\alpha$-(BEDT-TTF)$_2$KHg(SCN)$_4$ at the field perpendicular to the layers. With tilting the field, $G$ reduces, being determined by $B_\perp = B\cos\theta$, whereas the Pauli effect remains unchanged. This causes the transitions at low enough $\cos\theta$. With further increasing $\theta$, the transitions shift to lower fields, in agreement with the experiment.

Thus, the presented qualitative model seems to explain the physical origin of the multiple field-induced transitions in $\alpha$-(BEDT-TTF)$_2$KHg(SCN)$_4$ and their evolution with changing the field orientation. The real phase lines shown in Fig. 3 look somewhat more complicated than one would derive from this simple consideration. A more thorough theoretical analysis aimed to provide a quantitative description of the new phenomenon is in progress.

Concluding this section, we note that, by contrast to the "conventional" FISDW or FICDW cases which clearly require a strongly imperfect nesting of the FS ($t'_\perp > t'^*$), the new transitions can occur even at $t'_\perp < t'^*$. In particular, they can be observed in $\alpha$-(BEDT-TTF)$_2$KHg(SCN)$_4$ even at ambient pressure when the groundstate is a CDW. Another interesting point to be mentioned is that the quantum number $N$ of the CDW state *increases* with the field, in contrast to what is usually observed in known orbital quantization phenomena.

## 3. SUPERCONDUCTIVITY *VERSUS* CDW

In the $\alpha$-(BEDT-TTF)$_2M$Hg(SCN)$_4$ family only the $M$ = NH$_4$ salt is normal metallic and superconducting (SC) at ambient pressure with $T_c \cong 1.5$ K, the others, with $M$ = K, Tl and Rb, being in the CDW state. By applying a uniaxial strain in an appropriate crystallographic direction the $M$ = K and NH$_4$ salts have been shown to be switched, respectively from the CDW to SC state [17,18] and vice versa [17]. The hydrostatic pressure is known to suppress the CDW state in the K-salt, however, no superconductivity at high pressures has been reported until very recently [16]. In fact, the K-salt does become a superconductor at a quasi-hydrostatic pressure $P \geq P_c$ [16] but the maximum critical temperature, $T_c \leq$ 0.11 K is 20 times lower than that found in the uniaxial strain experiments [17,18]. It was argued [25] that a compression along $c$ and, to a lesser extent, along $b^*$ direction enhances the SC instability while a compression along the $a$-axis (corresponding to the 1D direction) leads to an opposite effect. More work is necessary in order to understand whether the very low $T_c$ observed at the quasi-hydrostatic pressure is simply a coincidental result of a simultaneous compression in all three directions or it contains a new interesting physics.

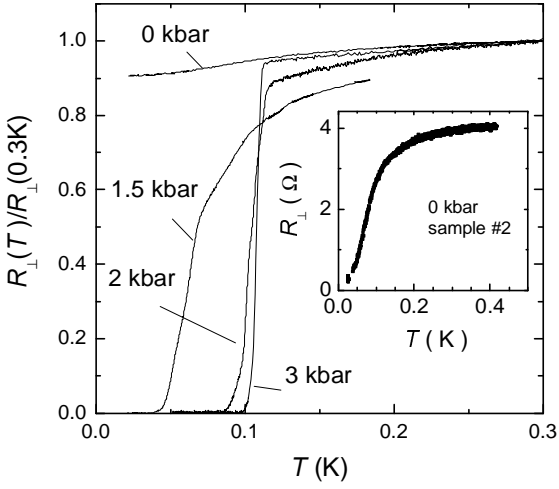

Figure 4. SC transitions in α-(BEDT-TTF)$_2$KHg(SCN)$_4$ at different pressures. Inset: ambient-pressure SC transition on another sample of the same compound.

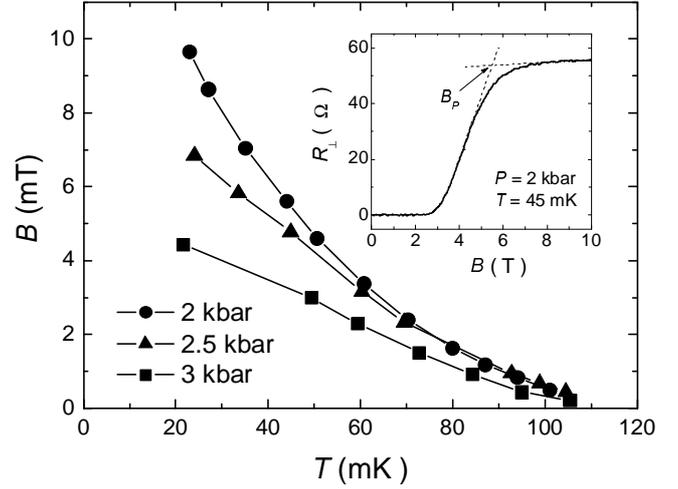

Figure 5. Temperature dependence of the critical field defined as shown in the inset, at pressures slightly below (2 kbar) and slightly above (2.5 and 3 kbar).

When the pressure is decreased below $P_c \approx 2.5$ kbar, the CDW is stabilized; however, the superconductivity does not disappear immediately. As shown in Fig. 4 (main panel), the resistive SC transition shifts down in temperature and broadens as pressure is lowered but even at $P = 0$ the resistance slightly decreases, indicating an incomplete transition to the SC state. The latter observation is in agreement with the previous finding by Ito et al. [26]. It should be noted that the shape and magnitude of the SC transition below $P_c$ are strongly sample-dependent. To demonstrate this, an example of a zero-pressure transition on another sample is presented in the inset in Fig. 4. The resistance of this sample is nearly zero at the lowest temperature. This, however, does not mean that the whole sample is in the SC state: d.c. magnetization measurements performed on the same sample using the SQUID technique have not shown any Meissner signal down to below 10 mK. Therefore the zero resistance is most likely achieved via a percolation network of thin SC paths.

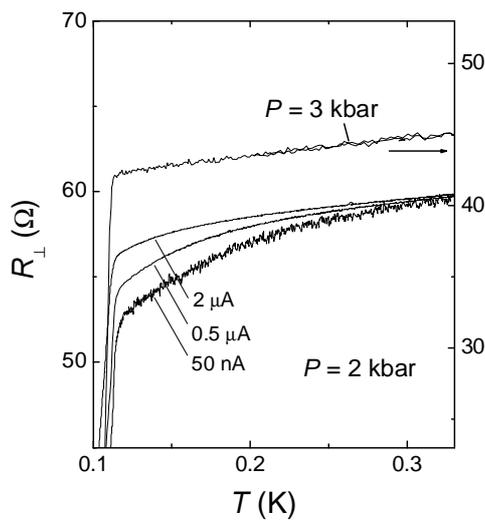

Figure 6. Onset of superconductivity at pressures above (3 kbar) and below (2 kbar) $P_c$. A clear SC precursor depending on current is seen in the lower pressure curves.

This suggestion is corroborated by the behaviour of the critical field perpendicular to the conducting $ac$-plane. Fig. 5 presents the temperature dependence of the critical field, determined as shown in the inset, at pressures around $P_c$. At zero field, the critical temperature does not change by more than 10% within this pressure interval. By contrast, the critical field shows a considerable enhancement at lowering the pressure, i.e. at entering the CDW region of the phase diagram. Moreover, the slope of the temperature dependence, linear above $P_c$, exhibits a clear positive curvature at the lowest pressure. We note that this behaviour remains the same independently of what point at the resistive transition, in either the temperature or field sweep, is chosen for the definition of the critical field. Similar results have been obtained on NbSe$_3$ in the region of coexisting CDW and SC phases [27] and in (TMTSF)$_2$PF$_6$ in which SDW and SC phases coexist in a narrow pressure range [28].

The presented results lead us to the conclusion that below $P_c$ the compound is in an inhomogeneous state in which tiny

SC fractions are included in a non-SC (actually CDW) matrix. The exact structure of this state is unknown at present; however, in analogy with NbSe$_3$, one could suppose that the SC is realized within boundaries between CDW domains where the CDW order parameter is suppressed [27].

So far, the described behaviour of α-(BEDT-TTF)$_2$KHg(SCN)$_4$ under pressure has been fully consistent with the generally accepted concept of a competition between the CDW and SC instabilities in low-dimensional conductors (see e.g. [29]): the fraction of the SC phase rapidly decreases upon entering the CDW region and $T_c$ shifts to lower temperatures. However, looking carefully at the onset of superconductivity, one notices that the resistance starts to deviate downwards from the normal metallic behaviour considerably above the temperature of the main resistive transition, as soon as the pressure is decreased below $P_c$. This is shown in Fig. 6 in which the $R(T)$ curves for a few pressures are presented in an enlarged scale. The resistance decrease at temperatures between ≈0.1 and 0.3 K seen in the $P = 2$ kbar curves is very sensitive to the transport current and can be suppressed by the field of ~10 mT. Therefore it most likely indicates that traces of superconductivity already exist at temperatures well above the maximum temperature of the bulk transition. The effect is either absent or greatly reduced at pressures above $P_c$. Further studies are needed to understand this apparent "enhancement of superconductivity" at entering the CDW state.

### Acknowledgements

The experiments in fields above 15 T were performed in Grenoble High Magnetic Field Laboratory. The work was supported in part by HPP Programme of EU, contracts HPRI-1999-CT-00030 and HPRI-CT-1999-40013, INTAS grant 01-0791 and RFBR-DFG No. 436 RUS 113/592/0-2.

### References

[1] Mori H., Tanaka S., Oshima M., Saito G., Mori T., Maruyama Y. and Inokuchi H., *Bull. Chem. Soc. Jpn*. **63** (1990) 2183; Rousseau R., Doublet M.-L., Canadell E., Shibaeva R.P., Khasanov S.S., Rozenberg L.P., Kushch N.D. and Yagubskii E.B., *J. Phys. I (Paris)* **6** (1996) 1527.
[2] Kartsovnik M.V., Kovalev A.E., Laukhin V.N., Schegolev I.F., Ito H., Ishiguro T., Kushch N.D., Mori H. and Saito G., *Synth. Met.* **70** (1995) 811.
[3] Foury-Leylekian P., Ravy S., Pouget J.-P. and Müller H., *Synth. Met.* **137** (2003) 1271.
[4] Basletic M., Korin-Hamzic B., Kartsovnik M.V., and Mueller H., *Synth. Met.* **120** (2001) 1021; Fujita T., Sasaki T., Yoneyama N., Kobayashi N. and Fukase T., *Synth. Met.* **120** (2001) 1077.
[5] Dora B., Virosztek A. and Maki K., *Phys. Rev. B* **65** (2002) 155119; *Physica B* **312-313** (2002) 571.
[6] Kartsovnik M.V. and Laukhin V.N., *J. Phys. (Paris) I* **6** (1996) 1753; Kartsovnik M.V., Biberacher W., E. Steep, Christ P., Andres K., Jansen A.G.M. and Müller H., *Synth. Met.* **86** (1997) 1933; Biskup N., Perenboom J.A.A.J., Brooks J.S. and Qualls J.S., *Solid State Commun.* **107** (1998) 503; Proust C., Audouard A., Kovalev A., Vignolles D., Kartsovnik M., Brossard L. and Kushch N., *Phys. Rev. B* **62** (2000) 2388.
[7] Christ P., Biberacher W., Kartsovnik M.V., Steep E., Balthes E., Weiss H. and Müller H., *Pis'ma Zh. Eksp. Teor. Fiz.* **71** (2000) 437 [*JETP Lett.* **71** (2000) 303].
[8] Harrison N., Balicas L., Brooks J.S. and Tokumoto M., *Phys. Rev. B* **62** (2000) 14212.
[9] Buzdin A.I. and Tugushev V.V., *Zh. Eksp. Teor. Fiz.* **85** (1983) 735 [*Sov. Phys. JETP* **85** (1983) 428].
[10] Zanchi D., Bjelis A. and Montambaux G., *Phys. Rev. B* **53** (1996) 1240.
[11] McKenzie R.H., cond-mat/9706235 (unpublished).
[12] Andres D., Kartsovnik M.V., Biberacher W., Weiss H., Balthes E., Müller H. and Kushch N.D., *Phys. Rev. B* **64** (2000) 161104(R).
[13] Lebed A.G., *Pis'ma Zh. Eksp. Teor. Fiz.* **78** (2003) 170.
[14] Ishiguro T., Yamaji K. and Saito G., Organic Superconductors (Springer-Verlag, Berlin, 1998).

[15] Chaikin P.M., *J. Phys. (Paris) I* **6** (1996) 1875.
[16] Andres D., Kartsovnik M.V., Biberacher W., Neumaier K. and Müller H. *J. Phys. IV France* **12** (2002) Pr9-87.
[17] Maesato M., Kaga Y., Kondo R. and Kagoshima S., *Phys. Rev. B* **64** (2001) 155104.
[18] Campos C.E., Brooks J.S., van Bentum P.J.M., Perenboom J.A.A.J., Klepper S.J., Sandhu P.S., Valfells S., Tanaka Y., Kinoshita T., Kinoshita N., Tokumoto M. and Anzai H., *Phys. Rev. B* **52** (1995) 7014.
[19] Kovalev A.E., Hill S. and Qualls J.S., *Phys. Rev. B* **66** (2002) 134513.
[20] Hasegawa Y. and Fukuyama H., *J. Phys. Soc. Jpn.* **55** (1986) 3978.
[21] Christ P., Biberacher W., Müller H., Andres K., Steep E. and Jansen A.G.M., *Synth. Met.* **70** (1995) 823; Christ P., Biberacher W., Jansen A.G.M., Kartsovnik M.V., Kovalev A.E., Kushch N.D., Steep E. and Andres K., *Surf. Sci.* **361/362** (1996) 909.
[22] Qualls J.S., Balicas L., Brooks J.S., Harrison N., Montgomery L.K. and Tokumoto M., *Phys. Rev. B* **62** (2000) 10008.
[23] Kartsovnik M.V., Andres D., Biberacher W., Christ P., Steep E., Balthes E., Weiss H., Müller H. and Kushch N.D., *Synth. Met.* **120** (2001) 687.
[24] Andres D., Kartsovnik M.V., Grigoriev P.D., Biberacher W. and Müller H., cond-mat/0305100; Phys. Rev. B **68** (2003) 201101(R).
[25] Kondo R., Kagoshima S. and Maesato M., *Phys. Rev. B* **67** (2003) 134519.
[26] Ito H., Kaneko H., Ishiguro T., Kono K., Horiuchi S., Komatsu T. and Saito G., *Solid State Commun.* **85** (1993) 1005; Ito H., Kartsovnik M. V., Ishimoto H., Kono K., Mori H., Kushch N. D., Saito G., Ishiguro T. and Tanaka S., *Synth. Met.* **70** (1995) 899.
[27] Briggs A., Monceau P., Nunez-Regueiro M., Ribault M. and Richard J., *J. Phys. (Paris)* **42** (1981) 1453.
[28] Lee I.J., Chaikin P.M. and Naughton M.J., *Phys. Rev. Lett.* **88** (2002) 207002.
[29] Gabovich A. M., Voitenko A. I. and Ausloos M., *Phys. Rep.* **367** (2002) 583.